\documentclass{article}

\usepackage{arxiv}
\usepackage[numbers,sort&compress]{natbib} 
\usepackage[utf8]{inputenc} 
\usepackage[T1]{fontenc}    
\usepackage{hyperref}       
\usepackage{url}            
\usepackage{booktabs}       
\usepackage{amsfonts}       
\usepackage{nicefrac}       
\usepackage{microtype}      
\usepackage{lipsum}
\usepackage{graphicx}
\graphicspath{ {./images/} }
\usepackage{setspace}       

\title{The Risks of Industry Influence in Tech Research}

\begin{document}
\maketitle

\begin{center}
{\large
Joseph Bak\textendash Coleman$^{a,b,c,d,e,f,1}$,\;
Cailin O'Connor$^{g,j}$,\;
Carl Bergstrom$^{a,i}$,\;
Jevin West$^{h,i}$
\par}
\vspace{0.75em}
\begin{onehalfspacing}
\small
$^{a}$Department of Biology, University of Washington, Seattle, WA, USA\\
$^{b}$Centre for the Advanced Study of Collective Behavior, University of Konstanz, Konstanz, Germany\\
$^{c}$Department of Collective Behavior, Max Planck Institute of Animal Behavior, Konstanz, Germany\\
$^{d}$Department of Biology, University of Konstanz, Konstanz, Germany\\
$^{e}$Santa Fe Institute, Santa Fe, NM, USA\\
$^{f}$Berkman Klein Center, Harvard University, Cambridge, MA, USA\\
$^{g}$Department of Logic and Philosophy of Science, University of California Irvine, Irvine, CA, USA\\
$^{h}$Information School, University of Washington, Seattle, WA, USA\\
$^{i}$Center for an Informed Public, University of Washington, Seattle, WA, USA\\
$^{j}$Center for Socially Engaged Philosophy, University of California Irvine, Irvine, CA, USA
\end{onehalfspacing}

\vspace{0.5em}
\small $^{1}$\textbf{Corresponding author:} \texttt{jbakcoleman@gmail.com}
\end{center}

\begin{abstract}
Emerging information technologies like social media, search engines, and AI can have a broad impact on public health, political institutions, social dynamics, and the natural world. It is critical to develop a scientific understanding of these impacts to inform evidence-based technology policy that minimizes harm and maximizes benefits. Unlike most other global-scale scientific challenges, however, the data necessary for scientific progress are generated and controlled by the same industry that might be subject to evidence-based regulation. Moreover, technology companies historically have been, and continue to be, a major source of funding for this field. These asymmetries in information and funding raise significant concerns about the potential for undue industry influence on the scientific record. In this Perspective, we explore how technology companies can influence our scientific understanding of their products. We argue that science faces unique challenges in the context of technology research that will require strengthening existing safeguards and constructing wholly new ones. 
\end{abstract}

\section*{Introduction}
Digital information technologies are reshaping our world at an unprecedented pace \cite{Bak-Coleman2021StewardshipBehavior}. The swift integration of these technologies into our daily lives has ignited concerns about risks to our social structures, individual well-being, public health, economic stability, democratic function, and ecological sustainability \cite{Banker2019AlgorithmWell-Being:,Haff2014TechnologyWell-being,Kross2021SocialSteps, Zarocostas2020HowInfodemic, Cinelli2020TheInfodemicb, OConnor2019TheSpread}. In response, regulators worldwide have started to develop mitigating policy based on relatively scant evidence. For these initiatives to succeed, it is crucial that the scientific community fosters a rich and impartial understanding of technology's impacts that can serve as a bedrock for evidence-based policy.

Unfortunately, the incentives of technology industries are often misaligned with broader societal well-being. Scientific evidence about an industry's impacts can motivate regulation that may reduce or eliminate profitability. When such conflicts arise, the affected industries may have incentives to obstruct or distort the gathering of evidence and its translation into policy \cite{Oreskes2010MerchantsWarming,Pinto2017ToTo}. The past century is replete with instances where industries under threat disrupted scientific progress, often with considerable success. 

To give just a few examples, automotive manufacturers have resisted safety regulations by shifting blame to the behavior of pedestrians and drivers \cite{Norton2007StreetJSTOR}. Tobacco and asbestos companies concealed and obfuscated scientific research on the harms of their products for decades \cite{Barlow2017History19001975,Pinto2017ToTo,Oreskes2010MerchantsWarming, proctor2012golden}. Fossil fuel companies continue to sow doubt and impede progress on addressing climate change \cite{Oreskes2010MerchantsWarming}.  The soda industry funds and shares distracting narratives about the impacts of sugar on health \citep{CarpenterSweet, greenhalgh2024soda, kearns2016sugar}. We should expect much the same regarding the detrimental effects of digital information technologies like social media, as well as search engines and AI.

Studying and regulating technology companies is particularly difficult because these industries restrict access to the data needed to inform policy \cite{Bruns2019AfterResearch}. Moreover, these companies employ vast teams of internal researchers who operate with limited external scrutiny, with no obligations to share their research, and with research budgets that dwarf those of even the best-funded academic labs. Furthermore, norms regarding the disclosure of industry funding in technology studies are lenient, granting companies significant discretion in shaping research directions within the academy.

These factors raise serious concerns about whether current scientific norms and institutions are sufficiently robust to resist industry influence in this growing area of research. In this perspective, we argue that existing safeguards are inadequate and that industry-driven manipulation of scientific research into information technologies is already well underway. If unchecked, we believe that the accumulation of the necessary evidence to underpin effective policy will be compromised. Indeed, it already has been. We believe these harms are largely unexplored and unacknowledged by those in the relevant research areas, so this perspective is critically important in promoting high quality research in this area.  We outline strategies that publishers, funders, scientists, and governments can adopt to protect and advance independent research into emerging information technologies.

\section*{Mechanisms of Industry Influence}
Policy that poses a threat to an industry's profits should be enacted when there is clear evidence of harm and a lack of good policy alternatives \cite{Rosenstock2011AttacksPolicy}. Reaching this point can take time, meaning such policy tends to be reactive rather than proactive. The process typically begins with informal observation or theorizing about potential harms, directing attention toward a potential problem. Researchers then need to establish evidence of clear causal links between an industry's conduct and undesirable outcomes. Finally, solutions must be proposed, and evidence gathered regarding their practicality and cost-effectiveness. Each link in this chain presents opportunities for industry to exert influence, and they do so in myriad ways. Many of these techniques are subtle and do not involve fraud or other overt violations of scientific norms, allowing industry to avoid risks of public censure and regulation \cite{Oreskes2010MerchantsWarming, OConnor2019TheSpread}.

\begin{figure*}[!ht]
\centering
\includegraphics[width=\linewidth]{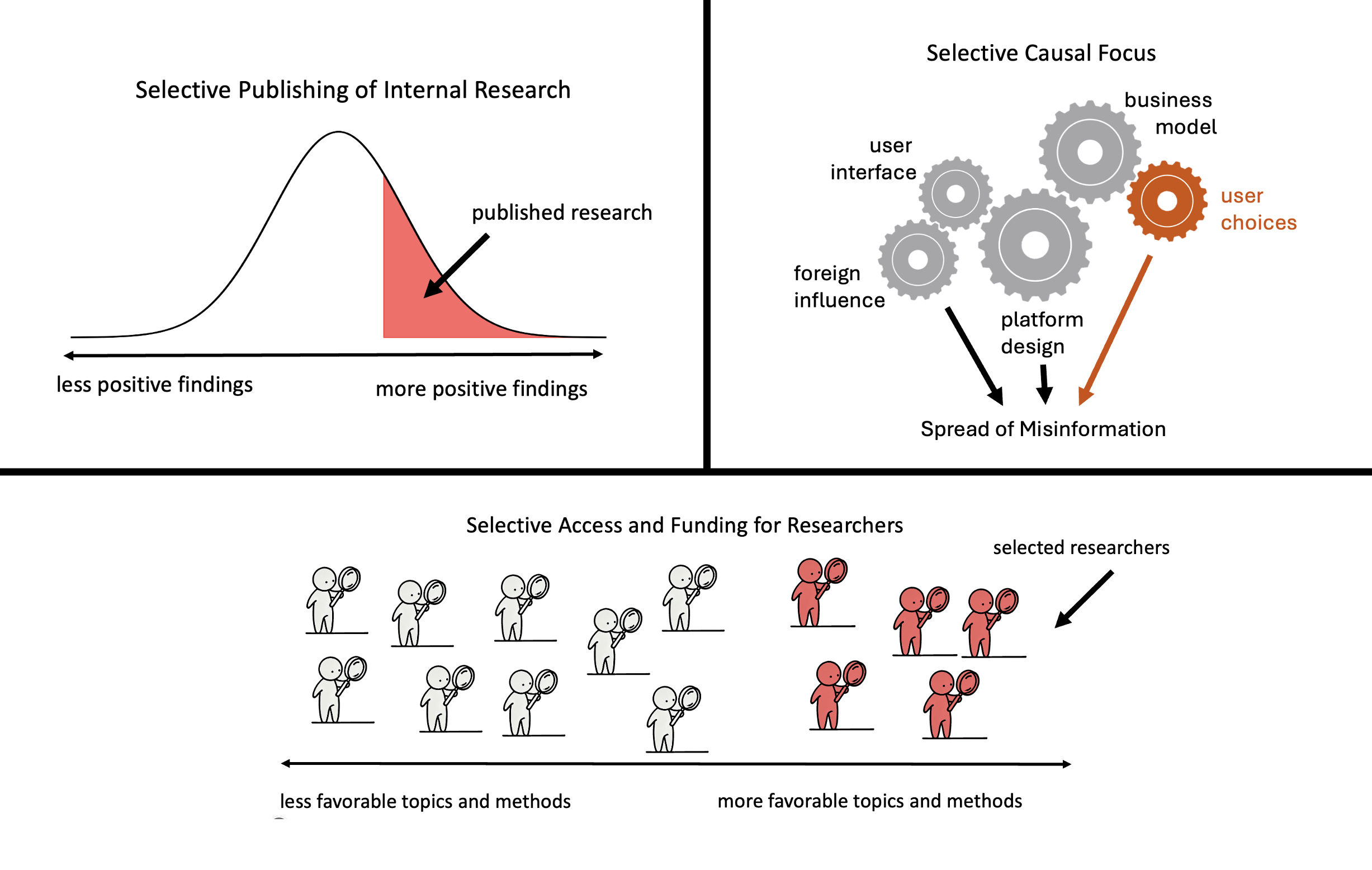}
\caption{Mechanisms of industry influence. Selective publishing, selective causal focus, and both funding of and promotion of selected researchers are all potential mechanisms for influencing research programs, questions, results, interpretations, and eventually policy. Importantly, none of these mechanisms typically involve fraud or other legal violations.}
\label{fig:fig1}
\end{figure*}

\subsection*{Burying Internal Research}
Perhaps the simplest, and least detectable, way in which industries impede the accumulation of evidence is by suppressing internal research that proves inconvenient. Companies are often uniquely positioned to detect harms caused by their product. Proximity to the product, involvement in its production process, and engagement with consumers provide them with relevant domain knowledge, privileged access to pertinent data, and strong incentives to address potential public relations or regulatory challenges proactively. Companies engaging in due diligence and product design will routinely uncover issues with their goods and services.  For example, leaked internal documents show that Meta staffers grew concerned about children lying about their ages to use VR products designed for older people as early as 2017, and have continued to raise red flags about child safety since \cite{swaine2025Meta}.

Sometimes, these problems are minimal or are easily remedied. Alternatively, the company may choose to go public with the issue and pay some survivable cost, such as initiating a recall. Yet in other cases, there may be no effective or financially desirable way to mitigate harm---and knowledge of such harm may even pose an existential threat to the industry.  The carcinogenic properties of asbestos provide one example, where companies were among the first to notice the impacts of asbestos on lungs as cases emerged among employees \cite{Barlow2017History19001975}. Similarly, in the late 1970s, Exxon-Mobil produced compelling evidence of climate change, alongside remarkably accurate predictions of carbon emissions and global temperature rise \cite{Supran2023AssessingProjections}. In each of these cases, the discovery had no profit-preserving remedy, and the industries chose not to disclose their findings. The result was decades of delay, profiting the companies at a massive social cost of inaction \cite{Barlow2017History19001975, Oreskes2010MerchantsWarming, Supran2023AssessingProjections}. Invariably, the severity of harm only become clear over time as independent research accumulated. More generally, if a company has reason even to suspect that such research may reveal existential threats, they have incentive to control or suppress it.

Unlike other industries that must decide to gather and analyze data on the impact of their products, the tech industry is unusual in that many of their products implicitly collect and analyze data---particularly about experiences and impacts on users \cite{Zuboff2023TheCapitalism}. Extracting insights from this data is an integral part of daily operations, essential for innovation and maximizing profits. In this regard, they may be more likely than most industries to identify harm in the course of routine operations. For example, content moderation undoubtedly provides direct insight into the types and frequencies of problematic posts that appear on a social media platform.

Such harms can easily remain concealed, as tech industry scientists are generally bound by extensive Non-Disclosure Agreements (NDAs) and require approval for their work to be published.  It is as if, in the case of climate change, oil and gas giants were the only ones with access to data on how fossil fuels impact climate, and had convincing legal reasons to keep it private.  As one might expect, these firewalls have proven effective at delaying disclosure. 

There have already been a number of leaks revealing suppressed internal research into harms caused by the tech industry.  These are likely indicative of a much larger volume of suppressed research. As noted above, four Meta staffers submitted a cache of internal documents to the US congress supporting their allegations that the company hid research about safety risks to children on their apps and VR devices \cite{swaine2025Meta}.  They report that Meta deleted and edited evidence of such harms gathered in the research process, and also that Meta lawyers advised researchers not to gather sensitive data, including about children using VR devices, to avoid regulatory concerns. In response, Meta has claimed that any such deletions would have been to protect the privacy of minors.  

Earlier documents leaked by Frances Haugen likewise revealed extensive internal research into Meta's platform impacts, including harmful effects such as Instagram's impact on teenage mental health \cite{Allyn20214NPR, milmo2021facebook}. Similarly, former Google employee Timnit Gebru was fired after refusing to retract published work on issues with large language models, now widely deployed \cite{Bender2021OnBig}. Together, these examples serve as evidence that tech companies routinely detect harm and choose not to disclose their findings, sanction employees who raise concerns, and even willfully avoid producing internal research on suspected harms.

\subsection*{Selective Publishing}
The flip side of undisclosed research is that companies can selectively publish research suggesting benefits or minimizing harms from their product. Companies can engage in internal review before research is made public, meaning work demonstrating benefits will tend to have an easier time finding its way into the pages of a scientific journal (Fig. \ref{fig:fig1}). This is a time-tested approach. Industry-published tobacco research, for example, painted a much rosier picture of its impacts compared to independent research \cite{Schick2005Scientific1929-1972,Oreskes2010MerchantsWarming}

With this technique, companies need not rely on fabricated results; instead, they can emphasize the real benefits of their products while neglecting to publish on its harms \cite{Jacquet2023TheBooks, freeborn2024industrial}.  This is typically possible since most products have both positive and negative social impacts.  They can also focus on research that by spurious chance shows no harm, even when such a harm exists \cite{weatherall2020beat}.  Relatedly, industry has a history of identifying and promoting favorable independent research that happens to be produced by academic scientists \cite{Oreskes2010MerchantsWarming,lewandowsky2019influence, weatherall2020beat,greenhalgh2024soda}. Together, these strategies create a skewed perception of the net benefits of some product while simultaneously avoiding fraud.

Meta, for example, has published work arguing that features of Facebook aid in grief following loss, improve longevity, benefit well-being, encourage reciprocity, reduce inequality, and mobilize voters in democracies \cite{Hobbs2016OnlineRisk, Hobbs2017ConnectiveFriend, Burke2010SocialWell-being, Kizilcec2018SocialGiving, Bond2012AMobilization, Chetty2022SocialConnectedness}. Each of these findings may well be true, but they represent an incomplete set of potential impacts and create a biased perception of net benefits and costs. Companies producing AI products likewise have promoted their potential and prospective benefits from improving crop yields and medical diagnoses to automating scientific discovery \cite{Dvijotham2023EnhancingClinicians, Kapetanovic2019FarmBeats:Agriculture, Yamada2025}.

Similarly, companies can selectively publish about interventions that worked to reduce their products' harms, creating a sense that they are actively progressing towards fixing issues whether or not this is true on balance. Social media companies have embraced this strategy, publishing promising results of platform interventions to combat hate speech and misinformation, even though these interventions do little to address the issues at scale \cite{Wojcik2022Birdwatch:Misinformation, Thomas2023DisruptingAudience, Chuai2024}. Historically, industry has over-emphasized the impacts of similar partially effective interventions like cigarette filters and plastic recycling \cite{freeborn2024industrial}.

Without access to internal records, it is very difficult to know when companies are selectively publishing favorable findings.  Recently, a paper published by Facebook employees indicated a substantial reduction in hate speech following a platform intervention that simultaneously banned multiple key accounts \cite{Thomas2023DisruptingAudience}.  This intervention was enacted in response to an uptick in hate speech during protests in 2020 \cite{Toraman2022BlackLivesMatterTwitter}. As hate speech across all platforms peaked and declined during this period, though, any interventions applied in response to the uptick could spuriously appear to indicate a long-term decline.  Notably, hate speech has been a problem for Facebook since its inception. One would certainly expect, and hope, that they have tried multiple strategies in response and evaluated their effectiveness. To date, however, we are not aware of any published research by Facebook on ineffective or actively harmful interventions they have tried. It is impossible for independent researchers to know whether this was the platform's first attempt at curbing hate speech; or simply the first one that yielded positive results. In our appraisal, the latter seems more likely.  

\subsection*{Design Bias}

\begin{figure}
\centering
\includegraphics[width=.8\linewidth]{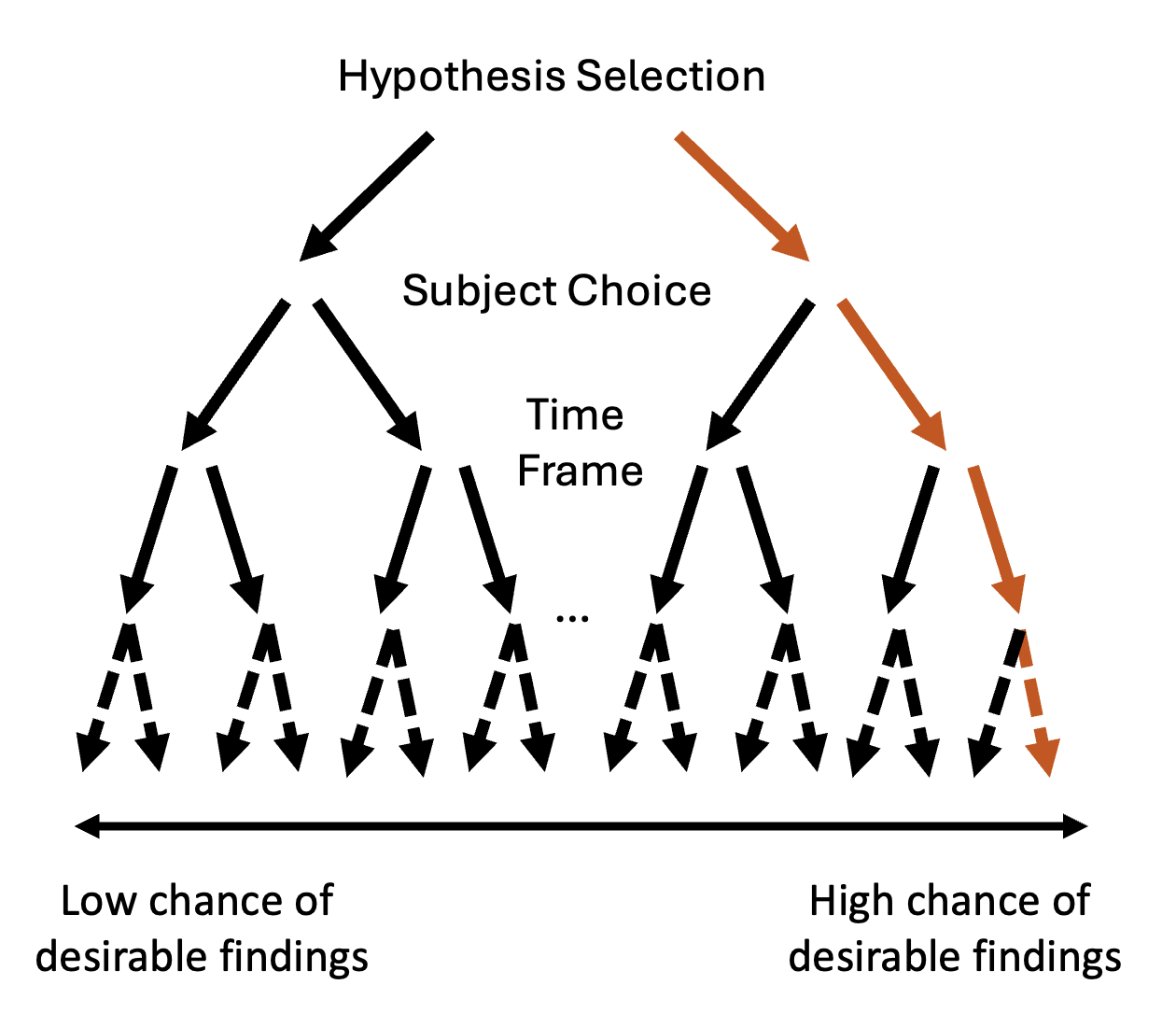}
\caption{Design bias. Throughout the process of study researchers can make choices more likely to yield desirable findings.}
\label{fig:fig2}
\end{figure}

Another worry about internal research---also demonstrated by the study just mentioned---relates to design bias.  In the process of designing a study, researchers must make myriad choices about exactly how to carry it out---what is the exact hypothesis, what exactly will be measured, who will the subjects be, how much data will be gathered, what will the time-frame be, for example. Even if a study is carried out transparently, these choices can have significant impacts on findings, and researchers aiming at a particular finding can make the choices most likely to yield that finding (see figure \ref{fig:fig2}).

Consider the ubiquitous choice of determining criteria for statistical significance and directionality of a test. Stricter criteria and adjustments for multiple comparisons will make null results more likely (and vice versa) \cite{gelman2013garden}. Meta published internal research that used the lenient significance criteria of $p<.1$ when measuring benefits in terms of well-being \cite{Burke2010SocialWell-being}. In the context of democratic outcomes, Meta researchers relied on conventional criteria  (i.e., 95\% C.I., $p<.05$) to establish benefits to turnout \cite{Hobbs2016OnlineRisk, Bond2012AMobilization}.  However, collaborations examining potentially deleterious platform effects on the 2020 election relied on more conservative corrections for multiple comparisons \cite{Guess2023HowCampaign, Nyhan2023LikemindedPolarizing,guess2023reshares, Allcott2024}. For example, a 2.6\% reduction in votes for Trump following withdrawal from the platform was ultimately deemed not significant following correction (Raw: $p=0.015$, Adjusted: $q=0.076$) \cite{Allcott2024}. By the criteria used to gauge benefits, this identical result would have led to the conclusion that Meta's platform did indeed alter voting patterns. 

Beyond statistical significance, researchers can choose study designs biased towards finding no evidence of harm. Most notably, while randomized controlled trials are a common causal workhorse in science, they can underestimate the potential for harm in the context of social media for several key reasons \cite{Bak-Coleman2025}. First, many experiments rely on interventions where subjects take short breaks from a platform, or from some features on the platform.  This is because participants actually unfamiliar with popular platforms are difficult to come by. Yet, we expect harms such as polarization, depression, and extremism to accrue over years of use.  It is unlikely they should fully revert in a withdrawal experiment lasting a few weeks \cite{Bak-Coleman2025}. In addition, it is difficult to practically isolate someone from social media given its widespread use---akin to being a non-smoker in an office full of smokers. Moreover, such studies can only provide insight into effects on individuals which may be small or undetectable even if societal-scale effects are pronounced \cite{Bak-Coleman2025}.  

Likewise, the choice of study endpoints can allow for additional bias. For example X (then Twitter) examined whether crowd-sourced fact-checking could combat the spread of misinformation \cite{Wojcik2022Birdwatch:Misinformation}. The authors focused on reduction in user engagement with labeled content, but measured such reductions only after a note had already been applied. The results showed meaningful effects on behavior. However, had the authors opted to measure the total effect on sharing---before and after labeling---they would have come to very different conclusion.  This is because over time content shows a natural decrease in engagement, and the measured reductions occurred after the bulk of sharing had already occurred \cite{Bak-Coleman2022CombiningMisinformation}. Indeed, subsequent research has shown that delays in applying notes substantially reduce their effectiveness \cite{Chuai2024}. 

These are just a few of the many ways design can shape study findings on tech products. Social media companies possess the capacity to iterate on study design in private. As such, even if published studies have designs that are publicly declared ahead of data collection and analysis, companies may have already selected for design features that increase chances of a desired outcome. Selectively reporting on research with biased design may also lead to proliferation of those designs in the literature since new studies tend to draw on older studies when making design choices---potentially biasing even independent future research. 

\subsection*{Selective Funding}
Although industry has significant control over internal research, a major drawback from a company's perspective is that this research may be viewed as less legitimate than independent research. Supporting academic researchers can increase the appearance of legitimacy while retaining some control over what research is conducted. Across domains, researchers that receive industry support have consistently produced work that better aligns with industry incentives. In medicine, for example, research funded by pharmaceutical or medical device manufacturers routinely reports higher effectiveness \cite{Lundh2017IndustryOutcome}. Tobacco-funded scientists were less likely to find links between second-hand smoke and lung cancer \cite{Schick2005Scientific1929-1972}. And systematic reviews by researchers with funding from the soda industry are much more likely to indicate no clear link between sugar and obesity \cite{Bes-Rastrollo2013FinancialReviews}. 

This does not imply that industry-funded researchers are engaging in ethically dubious, or industry designed, research. Science can be undermined when industry artificially props up otherwise earnest researchers whose work aligns with their incentives.   This phenomenon of ``industrial selection'' is made possible when, from a diverse pool of scientists, industry funds just those with favorable perspectives and practices, and otherwise leaves them to their own devices \cite{Holman2017ExperimentationSelection,Pinto2023EpistemicBias}. (See figure \ref{fig:fig1}.)  Given the feedback loop between funding, publishing, and career advancement, this strategy may also have long-term effects as funded scientists are promoted to positions of prestige, elevating their perspectives and practices \cite{Holman2017ExperimentationSelection, heyard2021value}. Notably, this strategy can be effective even if the funded scientists merely convince regulators that there is ongoing controversy in the relevant domain \cite{Porter2009Agnotology:Ignorance, Oreskes2010MerchantsWarming}.  

A related strategy funds and shares research specifically intended to distract from industrial products as a central cause of public harm \cite{proctor1995cancer, proctor2012golden, OConnor2019TheSpread}. (See figure \ref{fig:fig1}.) Coca Cola, for example, funds extensive research into the link between sedentary lifestyle and obesity \cite{greenhalgh2024soda, CarpenterSweet}. Tobacco similarly promoted research into the link between asbestos and lung cancer \cite{Oreskes2010MerchantsWarming}.  This type of ``industrial distraction" or ``distraction science" makes use of real, even important, research that can nonetheless mislead about industry harms by suggesting other causes of those harms \cite{freeborn2024industrial, proctor1995cancer, proctor2012golden}. 

Industrial distraction is often used to shift the blame for a product's impacts from its design to its users. For most industrial products, public harms result from feedback between product design and user behavior. Thus, by emphasizing the causal role of the users in these harms, industry can de-emphasize the role of the product itself. For example, fossil fuel companies sponsored work on carbon footprints, and firearm lobbies have emphasized mental health rather than gun access as the central causal factor in gun violence \cite{Walenta2021TheScoping, Higenbottam1989TheCigarettes., McGinty2014News1997-2012}.  

Tech companies have embraced this strategy by shifting research away from the design of their technology and towards the actions and behaviors of users.  Misinformation and polarization research over the past decade provides a compelling example. Meta (then Facebook) created a study which argued that users were more responsible than the algorithm for causing polarization \cite{Bakshy2015ExposureFacebook}. More recent, industry produced work has argued that detrimental impacts of social media platforms are in part due to the negative mindsets individuals have about the impacts of social media, rather than the media itself \cite{Ernala2022MindsetsWell-Being}. 

Studies of helpful interventions have likewise focused on users. Google, Facebook, and Twitter all funded or facilitated research on positive ``nudges'' which tend to involve various forms of labeling designed to deter users from sharing undesirable content \cite{Allen2021ScalingCrowds, Roozenbeek2021How2020, Wojcik2022Birdwatch:Misinformation, Allen2022BirdsProgram}. There has been particular interest in relying on users to generate labels, increasing the perception that users can and should be responsible for preventing the sharing of misinformation.

\subsection*{Inhibiting Independent Research}
Neither selective funding and publication nor burying internal research will reliably prevent independent researchers from gathering data regarding an industry's impacts. If compelling enough, such findings may lead to financially damaging policy. Companies need some other way to prevent this, and one powerful mechanism involves limiting the ability of independent researchers to gather such evidence.

Most social media companies provided researchers with extensive access to data via application programming interfaces (APIs) throughout the mid 2010s, but this practice was curtailed in the wake of the Cambridge Analytica scandal \cite{Bruns2019AfterResearch}. The restriction of such data has intensified in the intervening years, with most platforms currently offering little or no API access for research. Meta has eliminated its widely used CrowdTangle platform \cite{ortutay2024meta}, and X has eliminated affordable academic options for data access \cite{murtfeldt2024rip}. Researchers denied data access must rely on ``scraping''---collecting public-facing data through automated tools. Although there are no laws in the US against scraping, technological solutions to reduce scraping are commonplace and can make data gathering difficult \cite{amarikwa2023internet}.

For researchers that rely on API access as well as scraping, fear of losing API access may also reduce willingness to publish research using scraped data. Moreover, researchers who have gathered data directly from users have sometimes faced legal threats. Facebook sued the NYU Ad Observatory for collecting data without their authorization \cite{Sellars2020FacebooksLab}. Elon Musk sued the Center for Countering Digital Hate after they published a paper which used scraped data to show an uptick in hate speech following his acquisition of Twitter \cite{Allyn20214NPR}.  

Companies typically require researchers to adhere to a data use agreement if they wish to use data collected via APIs. While fairly standard, these instruments provide ample leverage for industry influence. For example, APIs typically require that researchers delete anything the company has itself  removed \cite{Lurie2023ComparingRequirements}. Although such restrictions are justified under the auspices of preserving privacy, academic research that systematically lacks access to policy-violating and extreme content will necessarily underestimate the abundance of harmful content on platforms. Moreover, nothing prevents a company from wielding these terms of service to prevent publication of research. In practice, such terms place sufficiently stringent demands on academic use as to place researchers in persistent legal jeopardy. For example, X sent some researchers notices requiring that they delete all existing data or pay fees of \$42,000 a month \cite{Calma2023TwitterResearch}. 

Even when companies do provide access to data, extensive delays or poor data quality can consume resources and frustrate projects. Researchers that joined a data-sharing program with Meta called Social Science One wound up waiting almost two years for the data to become available \cite{Allen2021ResearchData, silverman2019}. Worse yet, it was only discovered after several papers were produced that the dataset was missing nearly half of the users with identified political preferences. Although there is no indication these issues were intentional, lack of incentives to share data correctly and quickly still undermines independent research. As data requests become more prominent under the European Digital Services Act, this may be a recurring issue. 

Equally concerning is that researchers have no way of independently cross-checking data that social media companies provide. For example, the URL data provided by Meta to Social Science One could not be cross-checked or verified with any other data \cite{Hegelich2020blog}. In addition, citing privacy protections, Facebook restricted researchers' data access but provided less restricted data for free to private companies \cite{hegelich2020facebook}. 

Note that all this data is further limited in that it only provides insight into the behavior of users. What is missing is information about the platform, from its technological features (e.g., algorithms, spam detection) to its sociological ones (content moderation, monetization). For much of the past decade, independent researchers have been given no industry-provided tools for directly probing the function of algorithmic systems, despite the fact that such tools exist for in-house use \cite{Huszar2022AlgorithmicTwitter}. Likewise, data gathering APIs historically have provided some access to the types of data users share or engage with, but very little insight into what users see \cite{Bruns2019AfterResearch}. Similarly, researchers have not been provided tools for probing the functioning of recommendation systems to evaluate their impacts. As a consequence, research has been forced to focus disproportionately on user behavior rather than platform design, or interaction between these two elements, which, as noted, may shift perceptions of responsibility away from platforms. 

\subsection*{Selective Access and Performative Collaboration}
When tech companies do share data with independent researchers, they may do so selectively in order to promote industry messaging. We call this ``performative collaboration"---collaborations implemented with pragmatic industry goals in mind, rather than with a primary purpose of yielding high quality research.

For example, the long-standing lack of academic access to social media data was recently relaxed through a collaboration between Meta and academic researchers. A pair of academic leads, with existing ties to Meta, were selected to work alongside Meta researchers \cite{Wagner2023IndependencePermission}. They were free to select their academic team, provided members had previously engaged with Meta in their Social Science One project. All experimental designs had to be approved by Meta, raising the possibility of selection for design-biased research. The independent rapporteur charged with overseeing this referred to the process as ``independence by permission" \cite{Wagner2023IndependencePermission}.

This kind of selective access, besides leading to design bias, may have follow-on impacts on larger research communities.  Researchers selected to participate will tend to have privileged opportunities for career growth \cite{heyard2021value}. Top journals prefer to publish studies with large data sets, and these researchers have access to more data and more funding than most of their peers. It is similarly not uncommon for industry researchers to return to the academy, having published for years with unique access to data and resources. 

As these industry associated researchers rise to prominence, they often become gatekeepers---such as editors and peer reviewers---in top journals.  We are not suggesting they are engaging in any sort of misconduct, but rather that, in line with our discussion of industrial selection, researchers selected by industry because of their methods, topic choices, or approaches may end up disproportionately impacting the direction of technology research. Something similar happened in cardiovascular research when pharmaceuticals funded and promoted only scientists using the methods that tended to find anti-arrhythmic drugs were safe and effective.  These scientists often rose to disciplinary prominence on the basis of that funding \cite{Holman2017ExperimentationSelection}.

These performative collaborations can also have multiple, positive impacts for companies beyond their impacts on research. Conveying cooperativity and a sense of public concern, they can generate positive associations for the company. In this way, these collaborations can function similarly to green-washing, where companies publicize minimal environmental actions to create a positive public image \cite{de2020concepts}. As researchers in the field have commented, the Social Science One 'collaboration' following the Cambridge Analytica scandal was an attempt by Facebook to "gain positive news coverage and to reduce political pressure on the company" \cite{hegelich2020facebook}.  Throughout the history of industry influence on science, companies have likewise laundered the reputations of independent scientists in various ways, including by the simple association of academic science with their brand  \cite{Oreskes2010MerchantsWarming, OConnor2019TheSpread}.  Recently, for example, Coca Cola has been criticized for performative academic collaborations in exercise science that both allow them to control which topics get researched and also create the impression that Coke cares about public health \citep{greenhalgh2024soda, CarpenterSweet}.

An ongoing collaboration between the Center for Open Science and Meta reveals the degree to which Industry-Academic collaborations have shifted towards performative, data-contained, and design biased.\footnote{Details of the collaboration can be found here: https://www.cos.io/about/news/meta-partners-with-cos-to-share-data-to-study-well-being-topics} In this collaboration, four selected labs will be provided data about teen users' of the platform, but are prohibited from requesting the specific content users saw, engaged with, or shared. By contrast, Meta's internal research seems to have identified content-specific harms arising from exposure to content like that containing body standards, bullying, and unwanted sexual advances \cite{USSenateSubcommitteeonPrivacy2023,Wells2021}. For example, Meta found that teens with body image issues were exposed to feeds containing 10.5\% ``“eating disorder adjacent content,'' compared to 3.3\% for those that did not \cite{Horowitz2025}. Yet this data, known to Meta, is not being provided to independent researchers. As a result, any impacts of exposure to harmful content cannot be revealed by the collaboration, resulting in design bias away from plausible sources of harm.

\section*{Insufficiency of existing norms and institutions}
\subsection*{Publishing requirements}
Past conflicts between science and industry have led to the development of systems intended to identify and reduce industry influence. Clinical trial registries were developed and strengthened in medicine in response to concerns over misleading presentations of biomedical research \cite{Ioannidis2005ContradictedResearch}. By contrast, journals publishing technological research do not typically require trial registries, even when experimentally evaluating their products' impacts on health and well-being. 

When registries are not deemed necessary, academic journals and institutions nonetheless require scientists to have their work approved by Institutional Review Boards (IRBs) which function to weigh potential risks to study subjects against benefits to science and society. This process provides important safeguards, constraining what science can be conducted. However, for reasons that are unclear to us, journals routinely waive IRB approval as a requirement for industry-produced research. For example, this has been the case for  experimental research aiming to alter emotions, voting patterns, and content exposure \cite{Huszar2022AlgorithmicTwitter, Kramer2014, Bond2012AMobilization, Thomas2023DisruptingAudience}. As a result, industry is able to produce and publish research that independent scientists ethically cannot.  Notice this asymmetry also prevents independent replication of industry findings.

Competing interests declarations are another important mechanism that enables readers to appropriately reduce trust in industry influenced work \cite{Schroter2004DoesTrial}. However, there is evidence that tech-funded researchers are regularly failing to disclose industry funding or affiliation. Meta's research blog identifies more than a thousand individuals to whom it has funded or awarded fellowships. (This includes only recipients of open awards---not of informal direct grants.) Yet, a search of OpenAlex indicates only 472 publications produced by independent academics acknowledging Facebook/Meta as a funding source. In writing this piece, we identified numerous instances where academics collaborating with or funded by industry failed to note these competing interests. In addition, in several cases, academics with substantial funding from a given company nonetheless served as editors and peer-reviews of research relevant to their funders. 

To be clear, this lack of transparency should not be understood as a deliberate attempt to obscure questionable financial relationships. Rather, it reflects existing norms in the field coupled with a general lack of awareness around the subtle ways industry can impact the progress of research. Indeed we found several instances in which the authors disclosed industry funding yet affirmatively declared no competing interests---clearly in conflict with many journal's policies on the matter. Our guess is that researchers often perceive their ties to industry as unimportant or irrelevant to their research choices, despite that in other areas of science comparable ties have historically had substantial impacts on research outcomes. 

In sum, current publishing requirements aimed at protecting the integrity of science are not always functioning to do so when it comes to tech research.

\subsection*{Open Science and related reforms}
Industry-academic collaborations have leaned on ``Open Science'' in recent years to address concerns over transparency \cite{Wagner2023IndependencePermission, Nosek2015PromotingCulture, Nosek2018TheRevolution}. A stated goal of open science is to improve transparency through sharing of data and materials as well as \textit{a priori} commitments to design and analysis through preregistration \cite{Nosek2015PromotingCulture}. Yet there are reasons to think these policies cannot fully protect tech research from industry influence. On the contrary, tech companies may sometimes leverage open science to their advantage while independent researchers remain constrained by its requirements \cite{Devezer2021TheReform, Pinto2020CommercialScience}. 

Preregistration, for example, involves creation of a publicly accessible, time-stamped document outlining study hypotheses and planned analyses. In 2020, a collaboration between Meta and academic scientists evaluated impacts of misleading and inflammatory content on democratic processes. They pointed to preregistrations performed as an assurance that they ``didn't hide any of the results".\footnote{See https://research.facebook.com/2020-election-research/} Yet it is well established that authors routinely deviate from their plans in ways that are not disclosed or difficult to identify \cite{Claesen2021ComparingStudies}.  Although deviations can be justifiable, each provides an opportunity to hide undesirable results. 

The preregistered collaboration just described involved many such deviations, with one study disclosing three dozen \cite{Nyhan2023LikemindedPolarizing}.  For example, it was later revealed Meta deviated by initiating 63 secret, temporary measures targeting harmful content on their platforms. These changes were found to reduce the presence of such content during the study period \cite{Guess2023HowCampaign}. By analogy, this amounts to installing temporary air filters on factories only while measuring their pollution levels. As this deviation and its impacts were not disclosed at the time of publication, it was not taken into account by reviewers and editors. 

Some industry research---including an ongoing collaboration between the Center for Open Science and Meta---relies on registered reports to reduce the potential for bias. These are similar to preregistration, but involve pre-review of the study, and a guarantee of publication regardless of results. But industry can (in principle) conduct an analysis and examine results before deciding whether to collaborate with academics to submit a registered report. Without independent supervision of the full research process it is impossible to know if registered reports are properly constraining industry research in these cases. Given the publication guarantee, any obfuscation that occurs during data collection and delivery will invariably get the stamp of ``peer-reviewed'' research.   

An additional worry is that independent researchers are restricted in their ability to preregister novel studies about online platforms. Some proponents of preregistration have argued that it is ineffective or inappropriate for existing, observed data \cite{Lakens31122024, Nosek2018TheRevolution}. However, given limited access to data, social media researchers routinely re-analyze existing datasets. Moreover, open science advocates describe work lacking preregistration as exploratory, less credible, and not useful for making inferences to any wider population \footnote{https://www.cos.io/initiatives/prereg}. Taken seriously, these standards could provide grounds for dismissing the vast majority of independent research as merely exploratory intrinsically as less rigorous than industry sponsored research.


A key tenet of Open Science involves sharing of data and code \cite{Nosek2015PromotingCulture}.  Industry affiliated researchers have sometimes publicly embrace this standard. Yet companies are free to quietly publish as much or as little of their data, code and materials as they want. Selective sharing is a concern.  In any case, if there is no independent appraisal of this entire pipeline from raw data gathering to publication, there is no reason to trust industry adherence to open data.  And, once again, these requirements may, paradoxically, disadvantage independent researchers who are often prohibited from publicly sharing API-collected data \cite{Lurie2023ComparingRequirements}. 




Open science norms surrounding large sample sizes and statistical power may, likewise, advantage industry \cite{Lakens2022SampleJustification}. Academic sample sizes are constrained by the costs of paying participants and by ethical considerations enforced by IRBs. By contrast, researchers in the tech industry have access to millions---sometimes even billions---of users who can be studied without their knowledge, much less consent or payment. For example, research on X's algorithmic impacts involved experiments conducted on millions of users without requiring IRB approval, participant consent, or payment \cite{Huszar2022AlgorithmicTwitter}. 

Other standards for rigorous science can also disadvantage independent tech research. For example, it is a common misunderstanding that to establish causal claims in science requires manipulative studies, which are generally only possible for platforms and platform collaborators, rather than independent researchers \cite{Deaton2018UnderstandingTrials, Bak-Coleman2025}. In reality, there are no shortage of techniques for establishing causality using observational data, and such approaches routinely and consistently uncover evidence of harm caused by social media \cite{Pearl2009CausalOverview,  Lorenz-Spreen2022ADemocracy, Abrahamsson2024Smartphone, 10.1257/aer.20211218,angrist2014mastering}. Nevertheless, confusion over this point leads journals to prioritize on-platform experimentation. As a result, high-profile reviews and experiments have dismissed non-experimental evidence suggesting harms \cite{Budak2024MisunderstandingMisinformation, national2023social,Nyhan2023LikemindedPolarizing,Guess2023HowCampaign}. 

Across these examples 1) industry can find ways around the standards of open science, and 2) well-intentioned reforms can become epistemic filters that select for industry-produced science. For industry, selective application of Open Science provides a means of creating a perception of unbiased, transparent work while retaining considerable potential to influence findings. For independent researchers, constraints posed by open science may be perceived as lack of transparency, rigor, or otherwise. For this reason, the otherwise laudable standards of open science may fail to preserve the integrity of tech research.

\section*{Proposed Solutions}

Despite the academy's best efforts, it is clear that most of the research findings about harmful impacts from tech remain hidden as trade secrets.  To protect the public, policies that bring this information to the light of day are needed.  In addition, we need policies to manage the tech industry's influence on academic research and the literature more broadly. 

To be clear we are not advocating an absolute firewall between industry and academia. Collaborations with industry can provide academic researchers with access to data that would otherwise be unavailable, and are also a powerful vehicle for transferring academic insights into real-world products and policies. Rather than throwing out the wheat with the chaff, we encourage more careful attention to  risks and trade-offs, alongside bulwarking norms and institutions against harmful influence in the field. 

A first step is to clarify and enforce existing norms within scientific institutions. One aspect of this would require ethical approval by an independent Institutional Review Board to publish any human subjects research in the technological sphere. Currently, industry-led experiments are publishable without such approval \cite{Thomas2023DisruptingAudience}. There is little reason to exempt industry from standard journal policies simply because their data arises from experiments conducted as a part of everyday business. Beyond the obvious ethical benefits, requiring IRB approval, perhaps alongside study registration, would provide insight into the proportion of attempted research that goes unpublished by industry. 

In addition, although most scientists are generally aware of the need to report financial conflicts of interest (e.g., direct payments and stocks), there appears to be variation across authors in declaring past funding, unique access to data, and collaboration as competing interests. Given apparent variation in disclosure norms, journals which have well-established policies may need to audit past research to identify undisclosed competing interests. Verification may be difficult at times, as there are few formal mechanisms in place to ensure that disclosures are accurate, or even to check if they are \textit{post hoc}. 

To avoid ambiguity going forward, journal policies should be adopted to ensure appropriate disclosures are clearly made by authors, reviewers, and editors, and to treat failures of disclosure as one might any other ethical issue in publishing. Such policies could be as simple as multiple prompts to researchers and reviewers for disclosure during the publication process and clearly communicating when such disclosures are required. As with other publication ethics lapses, journals can enforce consequences for authors (mandatory retraction), reviewers (removal from reviewer pool, or ethical investigation) and academic editors who fail to disclose conflicts of interests. To avoid risks of earnest mistakes, it may be prudent to create cross-journal databases that track COIs. 

More generally, it is important that reviewers, editors, and other stakeholders are, at the very least, aware of the biasing risks of industry-affiliated research and weigh those against more obvious strengths. In our experience, we find that most are simply not aware and industry ties are rarely considered when evaluating evidence. This is particularly important when the questions posed by a study could conceivably have outcomes whose discovery is misaligned with industry incentives. Reviewers and editors should explicitly consider whether the nature of industry collaborations could be biased by internal data collection practices or design. Reviewers and editors should pay particular attention when industry scientists could have had a ``first look'' at the results, enabling perhaps silent deviation, deletion or abandonment of the project conditional on the findings. Absent assurances in these domains, the risk of publishing a misleading study should be weighed against any insight it may provide. In less severe cases, editors and reviewers should consider whether claims surrounding the independence and verifiability provided by Open Science mechanisms are warranted. 

Of course, we should not rely on industry to be the sole producers of research on their impacts. For independent research, the constraints it operates under should be taken into account when assessing its contributions. There is no reason to prioritize potentially biased experiments over independent observational approaches to causality, nor large samples with high risk of bias over smaller independent samples \cite{Bak-Coleman2025}. Industry-imposed constraints on capacity for adhering to open science should not be penalized or viewed with suspicion. Likewise, brokers of scientific knowledge---those writing summary reports of literature on technology, or communicating key findings to policy makers---should take care to emphasize and value fully independent research when reporting their findings. For example, there is little reason for it not to be standard practice that tech journalists ask scientists if they have received industry funding or resources when working on a story related to the tech industry. Improving disclosure policy, as noted above, would be an invaluable tool for weighing independence as a strength of research in this domain---as we might consider any other methodological feature believed to substantially reduce potential bias.

Beyond leveraging existing institutions, leveling the playing field may require regulatory action. As noted, the last few years have seen an erosion of researcher access to tech data, jeopardizing the potential for even rudimentary independent research \cite{Bruns2019AfterResearch,Lurie2023ComparingRequirements, Wagner2023IndependencePermission}. This trajectory needs to reverse, and the European Union is leading the way with the Digital Services Act that sets the groundwork for mandatory researcher access to data, although its effectiveness remains to be seen. Nevertheless, this is an essential first step, and additional infrastructure will be needed to ensure these efforts are successful, and that data is accessible to researchers globally. 

Data access naturally comes alongside concerns about user privacy, and infrastructure will need to be built to balance trade-offs between privacy and access. Rather than relegating these choices to companies, institutional Review Boards can play an important role, ensuring that risks are justified by benefits when conducting research on digital data. At present, much of the research conducted in this space is considered exempt based on outdated policies of a pre-internet age. An update to the Belmont report, which outlines human subjects ethics, should explicitly consider risks and opportunities of research on digital data  \cite{anabo2019revisiting}. 

In some cases, additional privacy-minded practices may be required to reduce risks. These could include privacy-preserving datasets, or analysis in physical or virtual ``clean-rooms''---as used for U.S. census data---that ensure sensitive data is centrally maintained and not copied off premises \cite{jindal2024privacy}. Access to such clean rooms should be determined by independent approval boards free from competing interests, and not limited to those with whom industry wishes to associate. Although digital data will bring unique challenges, existing models of accessing sensitive biomedical data can provide a starting framework. 

In cases where there is evidence that internal research has been suppressed, government requirements for industry sharing of internal research may help \cite{wu2023should}.  Such requirements may state that after some time period, internal industry tech research must be publicly released. A worry is that such requirements might disincentivize industry research in the first place. But even legal requirements to simply disclose the nature of the experiments---if not their results---will help make clear what investigations are being performed, and what is not being publicly reported. There is clear precedent for such disclosures, and their benefits, in biomedicine and little reason it cannot be extended to social media companies conducting human subjects research. If industry is only willing to produce research on the condition that it be kept private, this should be a red flag. 

As noted a key limitation for independent researchers, compared to industry researchers, involves access to funding.  Meta reports spending \$43.8 billion on research and development in 2024 alone, with plans to increase that spending going forward.  If even a relatively small percentage of this goes to studying the human impacts of their technologies, that is still a sizable research budget for just one company.  On the other hand, during the entire four years of Biden administration, only \$267 million federal funidng was spent on grants studying misinformation and disinformation, some of these in the private sector. Even if independent researchers can access platform data, the resources they have available to study these platforms will never match those that tech companies can muster.  

Given this discrepancy, and the serious risk that emerging technologies pose for public health and democratic function, government bodies should prioritize funding this sort of research where possible.  As noted, there is a risk of feedback loops whereby industry-funded researchers are at an advantage when seeking funding from government sources \cite{Pinto2023EpistemicBias}. Here too, disclosure can facilitate directing research funds towards fully independent research. Unfortunately recent trends in US government funding are de-emphasizing or wholly eliminating streams of research on social media and especially misinformation, which may further exacerbate the worries raised throughout this perspective.

In some areas where there are worries about industry influence in science, independent scientists have worked together to produce regular reports about their subject matter.  The most notable of these is the UN Intergovernmental Panel on Climage Change (IPCC) which reports on  current consensus with respect to climate change.  A similar Intergovernmental Panel on Information Technology could protect the integrity of technology research both by producing an independent consensus document assessing new research in tech, and by pressuring tech companies to share data with independent researchers \cite{bak2023create}. Reports from the IPCC have been important to policy makers seeking to protect the public from climate change, and a similar report in the tech sphere might serve as an important tool for regulators seeking to reduce the harms of new information technologies. Given existing disclosure norms described above, it would be essential that a panel of this nature surfaces and considers competing interests among surveyed research and scientists nominated to contribute. 

Beyond the institutional solutions described above, awareness of the mechanisms of industry influence described above can facilitate informed decision making by academic researchers. As noted throughout, technology companies possess vast access to data, routinely run experiments, and employ large teams of scientists to make sense of the findings. For many research questions, the companies may be more than capable of conducting the research without academic involvement or may already have done so. In such cases, their goals for engaging in collaboration may be misaligned with genuine scientific curiosity. Researchers may instead be sent on performative, design-biased goose-chases or simply laundering the credibility of selected findings that align with industry interests. 

For researchers who decide to precede in industry-academic collaboration, it is critical to consider and take responsibility for risks of bias imposed onto ones' research. For example, do data-use agreements or other restrictions imposed by industry partners introduce design bias? Will the company you're engaging with have an opportunity to get a ``first peek'' at the results and decide whether to continue collaboration based on what they find? Moreover, if you are only being provided with processed data---what potential is there for data integrity issues arising during processing and how might you identify them? What data can you share, and will it allow for earnest independent critique your work? Questions like these are critical for determining whether to continue collaborating and, if so, whether there are any terms of the collaborative agreement need to be modified to ensure the final product is rigorous and unbiased.

\section*{Conclusion}
The distortion of scientific progress by industry is a common phenomena that has been documented across a wide range of disciplines and contexts. There is every reason to believe that research on digital technologies has been similarly impacted. The nature of research in these domains advantages industry in numerous ways; they have unique control of data, massive financial and scientific resources, and the capacity to wholly shield much of their product design from scientists' view. To protect inquiry in this area, and public interest more broadly, it is crucial that we find ways to mitigate industry influence in this space. 

Unfortunately, few if any safeguards currently exist to protect the legitimacy of technology research.  As we outline, there are practical policies that journals can easily implement that may greatly mitigate industry impact on the field.  Other reforms, such as requirements for industry sharing of data and internal research, require legislation and are thus less easy to implement, but may be necessary for reliable information environments in an online age.

\subsection*{Author Affiliations}

Joseph Bak-Coleman 
Department of Biology
University of Washington
Life Sciences Building 351800
Seattle, WA 98195-1800 USA
ORCID 0000-0002-7590-3824

Cailin O'Connor
Department of Logic and Philosophy of Science
University of California, Irvine
3151 Social Sciences Plaza A
Irvine, CA 92697 USA
ORCID 0000-0002-8351-2575

Carl T. Bergstrom
Department of Biology
University of Washington
Life Sciences Building 351800
Seattle, WA 98195-1800 USA
ORCID 0000-0002-2070-385X

Jevin D. West
Information School, Box 352840
University of Washington
Seattle, WA, 98195
ORCID 0000-0002-4118-0322

\subsection*{Acknowledgments}
A grant from Templeton World Charities (TWCF-2023-32581), awarded to J.B. and C.B  supported this research. The Knight Foundation to the Center for an Informed Public also supported elements of this research.
\subsection*{Delarations}
J.B. has served as a paid consultant to the United Nations on the impact of digital technologies.

\bibliographystyle{unsrt}   
\bibliography{references}

\end{document}